%% file: paper.tex
\documentclass{article} 
\usepackage{iclr2026_conference,times}

\input{math_commands.tex}

\usepackage{hyperref}
\usepackage{xurl} 
\usepackage{booktabs}
\usepackage{graphicx}
\usepackage{float}

\title{SPECA: Specification-to-Checklist Agentic Auditing for Multi-Implementation Systems — A Case Study on Ethereum Clients}

\author{
Masato Kamba$^{1}$, Akiyoshi Sannai$^{2,1,3}$ \\
$^{1}$Nyx Foundation, Tokyo, Japan, $^{2}$Kyoto University, Kyoto, Japan, \\ $^{3}$Research and Development Center for Large Language Models, National Institute of Informatics, Tokyo, Japan \\
\texttt{masato.kamba@nyx.foundation, sannai.akiyoshi.7z@kyoto-u.ac.jp}
}

\iclrfinalcopy 
\begin{document}
\maketitle

\begin{abstract}
\input{sections/abstract}
\end{abstract}

\input{sections/introduction}

\input{sections/related_work}

\input{sections/methodology}

\input{sections/evaluation}

\input{sections/conclusion}

\bibliography{references}
\bibliographystyle{iclr2026_conference}

\input{sections/appendix}

\end{document}

%% file: math_commands.tex
\usepackage{amsmath,amsfonts,bm}

\def\eqref#1{equation~\ref{#1}}

\def\1{\bm{1}}

\DeclareMathAlphabet{\mathsfit}{\encodingdefault}{\sfdefault}{m}{sl}
\SetMathAlphabet{\mathsfit}{bold}{\encodingdefault}{\sfdefault}{bx}{n}



%% file: sections/abstract.tex
Multi-implementation systems (e.g., protocols with independent clients) are increasingly audited against natural-language specifications. Differential testing scales well when implementations disagree, but it provides little signal when all implementations converge on the same incorrect interpretation of an ambiguous requirement. We present SPECA, a \textbf{SPE}cification-to-\textbf{C}hecklist \textbf{A}uditing framework that turns normative requirements into property-based checklists, maps them to implementation locations, and supports cross-implementation reuse.

We instantiate SPECA in an in-the-wild security audit contest for the Ethereum Fusaka upgrade, covering 11 production clients. Across 54 submissions, 17 were judged valid by the contest organizers. Cross-implementation checks account for 76.5\% (13/17) of valid findings, suggesting that checklist-derived 1$\rightarrow$N reuse is a practical scaling mechanism in multi-implementation audits. To understand false positives, we manually coded the 37 invalid submissions and find that threat model misalignment explains 56.8\% (21/37): reports that rely on assumptions about trust boundaries or scope that contradict the audit's rules. We also characterize key gaps: we detected no High/Medium findings in the V1 deployment, with misses concentrated in specification details and implicit assumptions (57.1\%), timing and concurrency issues (28.6\%), and external library dependencies (14.3\%). Our improved agent, evaluated against the ground truth of a competitive audit, achieved a strict recall of 27.3\% on high-impact vulnerabilities. This performance places the agent in the top 4\% of human auditors, outperforming 49 out of 51 contestants in detecting critical issues. While this evidence comes from a single deployment, it highlights a general systems-security lesson: formalizing threat assumptions early is critical for reducing false positives and focusing agentic auditing effort. Our approach demonstrates significant efficiency gains; the agent-driven process distills the audit, allowing human experts to validate findings and finalize a submission in just 40 minutes on average.

%% file: sections/introduction.tex
\section{Introduction}

Natural-language specifications increasingly define the security-critical behavior of complex systems, from network protocols and distributed ledgers to databases and runtimes. In these ecosystems, independent implementations evolve quickly and are audited under tight time and cost constraints. A persistent bottleneck is translating ambiguous textual requirements into executable checks and then verifying those checks across multiple implementations. Traditional automation (tests, static analysis, fuzzing) can validate particular code paths, but the spec-to-code mapping remains expensive and error-prone.

This bottleneck is especially acute in multi-implementation settings. Differential testing and cross-client comparisons scale well when implementations disagree, but they provide little signal when all implementations converge on the same incorrect interpretation of an ambiguous requirement. We refer to this failure mode as a \emph{semantic blind spot}: when correlated misinterpretations cause implementations to converge on the same (incorrect) reading of an ambiguous requirement, leaving divergence-based signals weak or absent. Addressing semantic blind spots is a general systems-security problem: audit effort can be consumed by low-value reports unless the threat model (trust boundaries and scope) is made explicit early.

We argue that specification checklisting provides a stable and traceable foundation for security auditing, enabling agents to map normative requirements to concrete code checks and reuse those checks across implementations. Cross-implementation checks are one strategy that benefits from this structure: once a checklist item fails in one implementation, the same item can be evaluated elsewhere, catching shared misinterpretations that differential testing would miss.

We implement this insight in SPECA, a \textbf{SPE}cification-to-\textbf{C}hecklist \textbf{A}uditing framework that decomposes the audit process into reproducible phases with inspectable artifacts. We evaluate SPECA in a real-world multi-implementation audit contest (the Ethereum Fusaka upgrade), and in this deployment Strategy B (cross-implementation checks) represents 76.5\% of all valid findings (13 of 17 issues). Our analysis also finds that 56.8\% of invalid submissions are explained by threat model misalignment---mismatches between assumed attacker capabilities or scope and the contest rules---suggesting that improvement efforts should focus on formalizing assumptions rather than solely increasing model capability. Our central message is that checklisting the specification and reusing checks across implementations can reduce semantic blind spots caused by correlated misinterpretations---but only when the threat model (trust boundaries and scope) is made explicit.

We re-evaluated the improved V2 agent against the Sherlock "Cant-be-evil" ground truth: on consensus-layer issues rated Low-severity or higher, it achieved 27.3\% strict recall (3/11), found two High-severity issues missed by V1, and surpassed 49 of 51 auditors (96

Existing approaches to detecting specification-implementation misalignment fall into two categories: differential testing and formal verification. Differential testing tools like Fluffy~\citep{Yang2021} and LOKI~\citep{Ma2023} have proven effective at finding discrepancies between implementations. However, these techniques fundamentally rely on behavioral divergence as their oracle: when all implementations agree on incorrect behavior, the bug remains invisible. Formal verification, while capable of proving correctness against specifications, requires substantial upfront investment in formalizing natural-language specifications into machine-checkable formats---a cost that few projects can bear, particularly as specifications evolve rapidly. What is needed is an approach that can systematically audit implementations against natural-language specifications at scale, without requiring full formalization.

In summary, this paper makes the following contributions:

\begin{itemize}
    \item \emph{SPECA: an artifact-centric framework for multi-implementation security audits.} We present a specification-checklisting approach that decomposes auditing into traceable artifacts---normative requirement extraction, specification-to-implementation mapping, property-based checklists, and audit reports with proof traces---and describe how these artifacts support systematic auditing (\S\ref{sec:design}).

    \item \emph{Cross-implementation checklist reuse as a scalable audit mechanism.} We operationalize ``1$\rightarrow$N reuse'' as a first-class strategy: once a checklist item is validated (or a defect is found) in one implementation, analogous checks can be propagated to other implementations to scale coverage (\S\ref{sec:design}--\S\ref{sec:evaluation}).

    \item \emph{In-the-wild case study (Ethereum Fusaka audit contest).} We instantiate the framework in a real-world security audit contest for the Ethereum Fusaka upgrade (11 production clients). In this deployment, cross-implementation checks account for 76.5\% of valid findings (13/17), illustrating practical scaling benefits under real operational constraints (\S\ref{sec:evaluation}).

    \item \emph{Failure modes and a systems-security design lesson.} Manual coding of invalid reports shows that threat model misalignment explains 56.8\% (21/37) of false positives---assumptions about trust boundaries or scope that contradict the audit's rules---highlighting that formalizing threat assumptions early is critical for reducing noise and focusing agentic audit effort. Missed High/Medium issues cluster in specification details and implicit assumptions (57.1\%), timing and concurrency issues (28.6\%), and external library dependencies (14.3\%), providing concrete targets for future enhancement (\S\ref{sec:evaluation}).

    \item \emph{Demonstration of Human Effort Reduction.} We quantify the manual effort involved, showing that our agent-driven pipeline significantly reduces the burden on human experts. The process allows a specialist to validate a core finding in approximately 10 minutes and refine a full PoC and report in another 30, demonstrating a substantial reduction in the manual effort required for complex audits.
\end{itemize}

%% file: sections/related_work.tex
\section{Related Work}
\label{sec:related}

We organize related work as three layers of a systems-security stack. (1) \textbf{Audit automation}: classic program-analysis and security-testing techniques (static analysis, fuzzing, symbolic execution, model checking) that provide scalable but often specification-agnostic signals. (2) \textbf{Divergence-based oracles in multi-implementation systems}: differential testing and differential fuzzing, which can scale across independent implementations but can become ineffective under \emph{correlated misinterpretations} of ambiguous requirements. (3) \textbf{LLM/agentic auditing}: approaches that use LLMs (often as tool-using agents) to align natural-language specifications with code and to generate structured audit artifacts. SPECA sits at the intersection of these layers: it reuses the scalability of multi-implementation settings while reducing reliance on divergence by anchoring checks in specification-derived artifacts, and it is designed to compose with classical analysis tools.

\subsection{Multi-Implementation Differential Testing}

Differential testing uses behavioral divergence as a bug oracle. For instance, Fluffy~\citep{Yang2021} applied this to Ethereum consensus, surfacing bugs via multi-transaction fuzzing. LOKI~\citep{Ma2023} extended this approach with state-aware fuzzing for consensus protocols. More recently, FORKY~\citep{Kim2025} introduced fork-state-aware differential fuzzing to specifically target fork-handling processes.

While highly effective at finding implementation divergences, these approaches share a fundamental limitation: differential testing cannot detect bugs when \emph{all implementations share the same misinterpretation} of the specification. Knight and Leveson's study~\citep{Knight1986} on N-version programming demonstrated that independently developed implementations often exhibit correlated failures rooted in specification ambiguity. Our work aims to mitigate this \emph{semantic blind spot} by grounding the audit in specification requirements rather than implementation consensus, thereby reducing dependence on divergence-based oracles that fail under correlated misinterpretations. Unlike differential fuzzing approaches that rely on behavioral divergence as an oracle, SPECA anchors correctness in the specification itself, enabling detection even under correlated misinterpretations.

\subsection{Specification-to-Implementation Alignment}

RFCAudit~\citep{Wei2025} is most closely related to our approach, proposing an agentic framework with separate indexing and detection agents to audit protocol implementations against RFC specifications. However, RFCAudit focuses on single-implementation audits, whereas SPECA targets multi-implementation environments where cross-implementation checks are evaluated as one of the strategies in our setting.

AUTOSPEC~\citep{Liu2025} tackles a complementary problem: synthesizing precise protocol specifications from natural language. Its emphasis on generating inspectable, traceable intermediate representations aligns with our artifact-centric design philosophy. Recent work on RFC-TCP/IP gap detection~\citep{RFCGap2025} demonstrates LLM-facilitated differential analysis between specifications and implementations, validating the feasibility of LLM-based specification understanding at scale. Unlike prior work that audits a single implementation or focuses on spec synthesis, SPECA targets multi-implementation security auditing and checklist reuse across clients.

\subsection{Autonomous Code Auditing Agents}

RepoAudit~\citep{RepoAudit2025} introduces agent memory, path-sensitive data-flow extraction, and validators for satisfiability checking. RepoAudit's validator-centric approach to reducing false positives differs from our emphasis on \emph{threat model formalization}: we find that the majority of false positives stem from mismatches between assumed and actual trust boundaries.

Multi-agent architectures have shown promise for vulnerability detection. VulAgent~\citep{VulAgent2025} employs hypothesis-validation with specialized agents. MAVUL~\citep{MAVUL2025} uses cross-role interaction among analyst, architect, and evaluator agents. While these approaches focus on deepening code understanding through agent collaboration, our work emphasizes \emph{specification as the source of truth} for what constitutes correct behavior. Unlike validator-centric or collaboration-centric agents, SPECA treats specification checklists and explicit threat model formalization as the primary levers for reducing false positives and scaling audits.

Neuro-symbolic and benchmarked approaches offer complementary angles. IRIS~\citep{IRIS2024} and QLPro~\citep{QLPro2025} couple LLM inference with static analysis (e.g., CodeQL) to synthesize checkers for CWE classes, while Agent Security Bench~\citep{ASB2025} and AgentDojo~\citep{AgentDojo2024} provide standardized evaluation suites for adversarial robustness. These works start from vulnerability-class taxonomies or benchmark scenarios; SPECA instead anchors on \emph{normative specification requirements} in multi-implementation settings, aiming to catch correlated misinterpretations that evade divergence-based oracles.

%% file: sections/methodology.tex
\section{System Design}
\label{sec:design}

This section presents SPECA, our specification-checklist-driven agentic auditing framework. We provide a high-level overview (\S\ref{sec:overview}), formalize the problem and threat model (\S\ref{sec:problem}), detail the two main phases (\S\ref{sec:phase1}--\S\ref{sec:phase2}), describe artifact generation (\S\ref{sec:artifacts}), present improvements based on failure analysis (\S\ref{sec:v2}), and cover implementation details (\S\ref{sec:impl}).

\subsection{Overview}
\label{sec:overview}

\begin{figure}[t]
\centering
\includegraphics[width=0.85\linewidth]{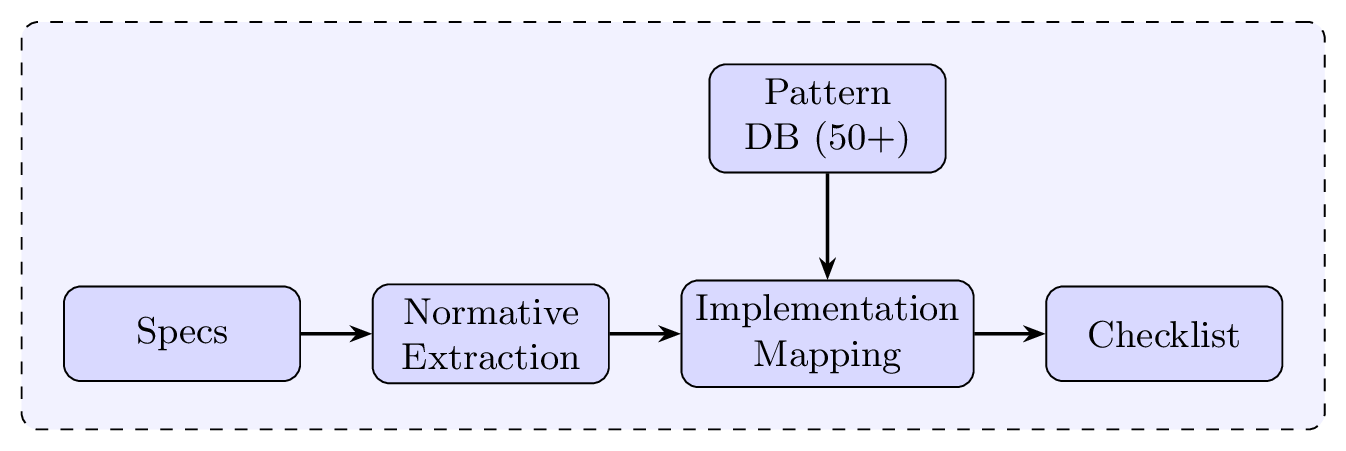}
\caption{Fig A: Checklist construction. Phase 1 extracts structured knowledge from specifications and produces a reusable checklist.}
\label{fig:architecture-a}
\end{figure}

\begin{figure}[t]
\centering
\includegraphics[width=0.95\linewidth]{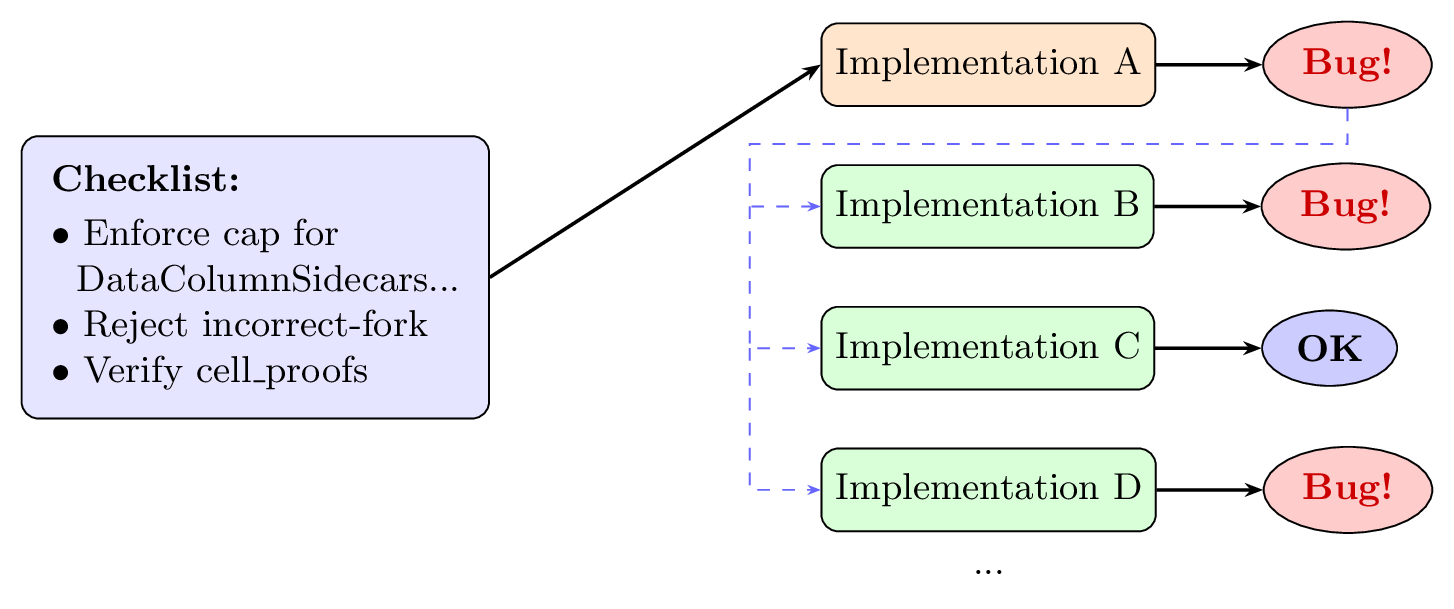}
\caption{Fig B: Strategy B cross-implementation checks. A checklist item identified on Client A is propagated across other clients, accounting for 76.5\% of valid findings in our deployment (\S\ref{sec:evaluation}).}
\label{fig:architecture-b}
\end{figure}

SPECA automates the expert security auditor's workflow through a two-phase pipeline: (1) \textbf{knowledge structuring}, which extracts and formalizes requirements from natural-language specifications, and (2) \textbf{systematic auditing}, which applies three complementary strategies to detect specification violations across multiple implementations. Figures~\ref{fig:architecture-a} and~\ref{fig:architecture-b} illustrate the checklist construction and cross-implementation auditing flow.

SPECA works by first building a structured representation of what the code \emph{should} do (from specifications), then systematically checking whether the code \emph{actually} does it (through implementation analysis). This checklist-driven approach helps mitigate semantic blind spots arising from correlated misinterpretations: because the checklist is anchored in the specification, it can flag shared misinterpretations that would not produce behavioral divergence.

Although cross-implementation checks are conceptually simple, our evaluation (\S\ref{sec:evaluation}) shows that they account for the majority of valid findings in practice, justifying their treatment as a first-class strategy.

\subsection{Problem Formalization and Threat Model}
\label{sec:problem}

\textbf{Problem Definition.} Given a natural-language specification $S$ consisting of normative requirements $\{r_1, r_2, ..., r_n\}$ and a set of implementations $\mathcal{I} = \{I_1, I_2, ..., I_k\}$ that claim conformance to $S$, the specification-checklist-based security auditing problem is to identify tuples $(r_i, I_j, v)$ where implementation $I_j$ violates requirement $r_i$ in manner $v$.

This formulation differs from differential testing, which identifies divergences $(I_j, I_l, x)$ where implementations produce different outputs on input $x$. Crucially, differential testing cannot detect violations when all implementations exhibit the same incorrect behavior.

\textbf{Threat Model.} We target vulnerabilities in protocol implementations that could lead to consensus failures, denial of service, or information disclosure. The specification documents are assumed correct (we detect implementation bugs, not specification bugs). Human verification is required for all findings. The threat model is treated as an explicit design parameter rather than an implicit assumption, as mismatches in assumed trust boundaries later emerge as the dominant source of false positives (\S\ref{sec:fp_analysis}).

\subsection{Phase 1: Knowledge Structuring}
\label{sec:phase1}

Phase 1 transforms unstructured specification documents into a queryable knowledge base.

\textbf{Specification Extraction.} The first component extracts normative requirements from specification documents. Given a specification document (e.g., an EIP markdown file), the extractor identifies statements containing modal verbs (MUST, SHOULD, MAY, SHALL) following RFC 2119 conventions and outputs structured requirements with unique traceability identifiers.

We use LLM-based extraction because specification documents use inconsistent formatting and natural language that resists rigid pattern matching. Each extracted requirement includes a unique ID (e.g., \texttt{EIP7594-CUSTODY-001}) and source location, enabling auditors to trace findings back to the original specification text.

\textbf{Pattern Database Construction.} The second component curates a database of known vulnerability patterns (50+ entries) relevant to the target domain, including boundary validation failures, guard condition omissions, state transition errors, cryptographic misuse, and resource management issues.

\textbf{Implementation Mapping.} The third component establishes correspondence between normative requirements and source code locations using a two-phase approach: keyword search to prune the search space, followed by semantic refinement using the LLM to ensure meaningful requirement-code correspondence.

\subsection{Phase 2: Systematic Auditing}
\label{sec:phase2}

Phase 2 applies three complementary strategies to detect specification violations.

\textbf{Strategy A: Specification-Based Static Audit.} This strategy directly checks whether each implementation satisfies the mapped normative requirements, examining presence (is there code addressing this requirement?), correctness (does the code correctly implement the requirement?), and completeness (are all conditions and edge cases handled?).

\textbf{Strategy B: Cross-Implementation Checks.} This is one of the strategies in SPECA. Strategies A and B share the same checklist derived from the specification; B evaluates the same checklist item across other implementations once a failure is observed.

The algorithm works as follows: (1) abstract a generalizable pattern from the initial finding, (2) find analogous locations in all other target implementations, (3) check each location for pattern matches, and (4) report cross-implementation findings for human verification.

\textbf{Why this strategy is effective in multi-implementation settings}: Violations often cluster around complex or ambiguous specification requirements. When one implementation team misinterprets a requirement, other teams---facing the same text---may make similar errors. Cross-implementation checks exploit this correlation structure by turning one finding into a systematic sweep across implementations. Representative examples are provided in Appendix Table~\ref{tab:cross_impl}.

\textbf{Strategy C: Dynamic Test Generation.} This strategy generates executable tests for boundary conditions. However, dynamic testing proved least effective in our evaluation (5.9\% of valid findings) because setting up test environments requires significant effort, and semantic blind spots can also reduce the usefulness of dynamic testing when the test oracle implicitly inherits the same correlated misinterpretations.

\subsection{Artifact Generation}
\label{sec:artifacts}

\begin{figure}[t]
\centering
\includegraphics[width=0.98\linewidth]{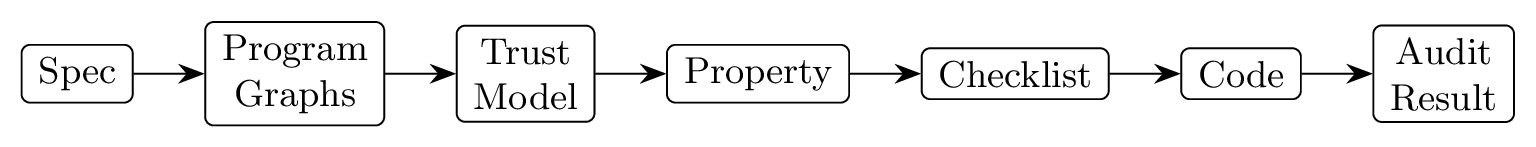}
\caption{SPECA V2 Agent Overview.}
\label{fig:v2-overview}
\end{figure}

Figure~\ref{fig:v2-overview} provides a concise overview of the SPECA V2 agent flow from specifications to audit outcomes.

\subsection{Improvements Based on Failure Analysis (V2)}
\label{sec:v2}

Analysis of false positives and missed findings in V1 revealed two primary areas for improvement:

\begin{enumerate}
    \item \textbf{Threat Model Misalignment} (56.8\% of false positives): The agent's assumptions about attacker capabilities did not match the actual system context.

    \item \textbf{Missed High-Severity Issues} (7 issues): Analysis \textbf{provides a clear roadmap for enhancement}: specification details and implicit assumptions (57.1\%), timing and concurrency issues (28.6\%), and external library dependencies (14.3\%).
\end{enumerate}

Based on this analysis, V2 introduces \textbf{TRUSTMODEL Formalization}: before auditing, V2 explicitly formalizes the threat model as a structured artifact specifying actor trust levels (TRUSTED, SEMI\_TRUSTED, UNTRUSTED) and boundary edges. This enables filtering findings that require capabilities outside the defined threat model.

To provide initial empirical evidence for V2, we conduct a post-hoc re-evaluation against the contest's ground-truth issue database. We re-run the V2 pipeline on the six CL clients and measure strict recall---counting only findings that identify the same bug class, code region, and root cause as a contest-validated issue. The results (Appendix~\ref{app:v2_reeval}) show that V2 achieves a strict issue recall of 27.3\% (3/11) on CL-related Low/Medium/High issues, successfully detecting 2 of 3 High-severity issues while missing issues that require deep stateful reasoning or client-specific implementation knowledge.

\subsection{Implementation}
\label{sec:impl}

We implement SPECA with a fixed LLM and custom prompts for each pipeline phase. We retain structured artifacts (specifications, maps, checklists, and reports) to support reproducibility and human verification.

%% file: sections/evaluation.tex
\section{Evaluation}
\label{sec:evaluation}

We evaluate SPECA as an in-the-wild case study of a broader systems-security problem: auditing multiple independent implementations of the same specification under tight operational constraints. We deploy SPECA in a production security audit contest for the Ethereum Fusaka upgrade (11 clients) and measure (i) how each audit strategy contributes to validated findings, (ii) which factors dominate false positives in practice, and (iii) which vulnerability classes remain challenging for a checklist-driven agentic approach.

\subsection{Research Questions}

RQ1: Which audit strategy contributes most to valid vulnerability discoveries in a multi-implementation environment?

RQ2: What are the primary root causes of false positives in LLM-assisted security auditing?

RQ3: What types of vulnerabilities does specification-checklist-driven agentic auditing fail to detect?

\subsection{Experimental Setup}

We evaluate SPECA in the Fusaka audit contest, a real-world multi-implementation security audit for the Ethereum Fusaka upgrade (Osaka EL + Fulu CL).

The contest covered 11 production clients and produced 366 total submissions from all participants, of which 101 were valid (27.6\% overall valid rate; full parameters in Appendix~\ref{tab:contest_params}).

Target Implementations. The contest covered 11 production clients: Nimbus, Lighthouse, Prysm, Teku, Lodestar, Grandine (consensus layer); Geth, Reth, Erigon, Nethermind, Besu (execution layer).

Scope and Severity. The contest scope is limited to Fusaka-specific code; known issues and out-of-scope findings are invalid. Non-informational findings must include an executable PoC at submission time. Severity is defined by network-impact thresholds (Critical $>$50\%, High $>$33\%, Medium $>$5\%, Low $>$0.01\%), and Informational issues are valid only if the client accepts and implements the change. These thresholds make Low/Informational findings operationally meaningful despite their label.

Attribution and Coding. We attribute each valid report to its primary discovery strategy as recorded in the submission: similar-issue-based analysis corresponds to cross-implementation checks, code analysis corresponds to specification-based static audit, and fuzz testing corresponds to dynamic test generation. We manually code invalid submissions into mutually exclusive categories and report the primary cause.

Human Effort. To quantify the manual effort involved in our process, we recorded the time spent on human-in-the-loop verification. On average, each generated report underwent an initial review of approximately 10 minutes to validate the core finding. For findings that proceeded to the next stage, an additional 30 minutes were spent on average to review, refine, and finalize the proof-of-concept (PoC) and the full submission report. This highlights the essential role of human expertise in validating and contextualizing the agent's outputs.

\subsection{Results: RQ1 -- Strategy Contribution Analysis}
\label{sec:rq1}

While most valid findings are classified as Informational under contest rules, such findings are operationally meaningful because they require client adoption to be considered valid (\S\ref{sec:evaluation}).

\begin{table}[t]
\caption{Overall submission outcomes. Our 31.5\% valid rate exceeds the contest average of 27.6\%.}
\label{tab:outcomes}
\centering
\begin{tabular}{lrr}
\toprule
\textbf{Metric} & \textbf{Count} & \textbf{Percentage} \\
\midrule
Total submissions & 54 & 100\% \\
Valid submissions & 17 & 31.5\% \\
Invalid submissions & 37 & 68.5\% \\
\bottomrule
\end{tabular}
\end{table}

Client Coverage: 9 of 11 clients (81.8\%)

Our approach achieved a 31.5\% valid rate (17/54), exceeding the contest average of 27.6\% (101/366). The 81.8\% client coverage indicates broad applicability across diverse codebases; the per-client breakdown is in Appendix Table~\ref{tab:clients}. However, all 17 valid findings from the V1 contest deployment were Low or Informational severity. We discovered none of the contest's 5 High-severity or 2 Medium-severity vulnerabilities in V1. This outcome distinguishes between two classes of challenges: issues rooted in implicit specification details, which are addressable by refining the SPECA methodology, and those stemming from complex stateful interactions, which highlight the current limitations of LLM agents. The analysis in Section~\ref{sec:rq3} further details this distinction.

\begin{table}[t]
\centering
\begin{minipage}[t]{0.52\textwidth}
\centering
\caption{Strategy attribution for valid findings. Cross-implementation checks represent 76.5\% of valid discoveries.}
\label{tab:strategy}
\begin{tabular}{lrr}
\toprule
\textbf{Strategy} & \textbf{Findings} & \textbf{\%} \\
\midrule
Cross-Implementation Checks & 13 & 76.5\% \\
Specification-based static audit & 3 & 17.6\% \\
Dynamic test generation & 1 & 5.9\% \\
\midrule
\textbf{Total} & \textbf{17} & \textbf{100\%} \\
\bottomrule
\end{tabular}
\end{minipage}%
\hfill
\begin{minipage}[t]{0.44\textwidth}
\centering
\caption{Severity distribution of valid findings.}
\label{tab:severity}
\begin{tabular}{lcccc}
\toprule
\textbf{Severity} & High & Med & Low & Info \\
\midrule
\textbf{Count} & 0 & 0 & 1 & 16 \\
\bottomrule
\end{tabular}
\end{minipage}
\end{table}

Cross-implementation checks contributed 76.5\% of valid findings in this deployment. This is consistent with the hypothesis that specification misinterpretations correlate across implementations: when one team misunderstands a requirement, others may make similar errors.

Summary for RQ1. In this deployment, cross-implementation checks represent the highest observed share of valid findings among the three strategies (76.5\%).

\subsection{Results: RQ2 -- False Positive Analysis}
\label{sec:fp_analysis}

We define categories as follows: threat model misalignment (assumptions about trust boundaries or scope that contradict contest rules), already known / duplicate (publicly known or previously reported issues), incorrect analysis (reasoning errors or misread code paths), and out of scope (findings not specific to the Fusaka upgrade).

The largest category (56.8\%) was threat model misalignment. Our prompts assumed execution layer (EL) responses were untrusted, while the contest rules treated EL as a trusted component. Under our coding scheme, this mismatch is associated with 21 invalid submissions.

Summary for RQ2. Threat model misalignment is the largest observed false-positive category (56.8\%), suggesting that assumption formalization is a high-leverage improvement target.

\subsection{Results: RQ3 -- Limitation Characterization}
\label{sec:rq3}

Notably, SPECA did not detect any High or Medium severity issues in this deployment. Our analysis of these misses distinguishes between categories that are addressable through methodological improvements and those that represent fundamental limitations of current-generation LLM agents.

We found none of the 7 High/Medium issues in the contest (Appendix Table~\ref{tab:miss}). The misses fall into three categories (Appendix Table~\ref{tab:failure} provides full detail).

\begin{table}[t]
\centering
\begin{minipage}[t]{0.48\textwidth}
\centering
\caption{False positive root cause distribution.}
\label{tab:fp}
\begin{tabular}{lrr}
\toprule
\textbf{Root Cause} & \textbf{Cnt} & \textbf{\%} \\
\midrule
Threat model misalignment & 21 & 56.8\% \\
Already known / duplicate & 8 & 21.6\% \\
Incorrect analysis & 5 & 13.5\% \\
Out of scope & 3 & 8.1\% \\
\midrule
\textbf{Total} & \textbf{37} & \textbf{100\%} \\
\bottomrule
\end{tabular}
\end{minipage}%
\hfill
\begin{minipage}[t]{0.48\textwidth}
\centering
\caption{Miss category summary for High/Medium issues.}
\label{tab:miss_categories}
\begin{tabular}{lrr}
\toprule
\textbf{Failure Category} & \textbf{Cnt} & \textbf{\%} \\
\midrule
Spec details \& implicit assumptions & 4 & 57.1\% \\
Timing and concurrency issues & 2 & 28.6\% \\
External library dependencies & 1 & 14.3\% \\
\bottomrule
\end{tabular}
\end{minipage}
\end{table}

\emph{Specification details and implicit assumptions (57.1\%---Addressable)}: The majority of missed issues (4/7) were due to subtle details in the specification that were overlooked, such as the exact fields to include in a cache key or nuanced input validation requirements. These gaps can be systematically narrowed by creating more granular checklists.

\emph{Timing and concurrency issues (28.6\%---Current LLM Limitation)}: Two issues were related to dynamic behavior that is difficult to capture with static analysis, such as metadata updates during fork transitions. These vulnerabilities depend on complex, state-dependent interactions that are difficult to reason about without a deep, stateful understanding of the system's runtime behavior, representing the frontier of current LLM agent capabilities.

\emph{External library dependencies (14.3\%---Addressable)}: One issue stemmed from a vulnerability in the `c-kzg-4844` library. This category is addressable by integrating automated tools like supply chain scanners into the audit pipeline.

Summary for RQ3. The analysis of missed High/Medium issues provides a critical distinction between improvable aspects of our methodology and the inherent limitations of today's LLM agents. While issues stemming from implicit requirements and external dependencies (71.4\%) can be addressed by enhancing checklist granularity and tool integration, complex timing and concurrency vulnerabilities (28.6\%) remain a fundamental challenge, highlighting a key area for future research in agentic reasoning and stateful analysis.

\subsection{Discussion}

\textbf{The Modern Landscape of Semantic Blind Spots.} Our findings offer a contemporary perspective on the nature of semantic blind spots in an era dominated by agentic coding. While classic correlated failures stemmed from ambiguous specification text, we hypothesize that the widespread use of a few powerful LLMs for development assistance introduces a new vector for shared vulnerabilities. If different teams leverage agents with similar underlying reasoning patterns, they may independently introduce homologous bugs, creating a deeper, more uniform blind spot that is even harder to detect with divergence-based methods.

This hypothesis provides a compelling explanation for our primary finding: \textbf{cross-implementation checks represent 76.5\% of valid findings}. In a world where agent-assisted development is common, a bug pattern found in one implementation is highly likely to exist in others, not just due to spec misinterpretation, but due to the convergent "thinking" of the agents themselves. The remarkable success of our 1$\rightarrow$N reuse strategy may therefore be a direct reflection of this modern reality, making it an increasingly critical tool for security auditing in the wild.

Our other findings are also interpreted through this lens. (1) Threat model misalignment remains the largest observed FP category (56.8\%); and (2) High/Medium issues were not detected in the V1 contest deployment (0\% detection rate), identifying concrete targets for future enhancement.

\textbf{Preliminary V2 Re-evaluation.} To assess whether V2's threat model formalization translates into improved detection capability, we conducted a post-hoc re-evaluation against the contest's ground-truth issue database (details in Appendix~\ref{app:v2_reeval}). Filtering to CL-client issues at Low severity or above (11 issues), V2 achieves a strict recall of 27.3\% (3/11), detecting 2 of 3 High-severity issues---including the proposer lookahead miscalculation and the c-kzg-4844 Fiat-Shamir weakness---and 1 of 6 Low-severity issues. The 8 undetected issues predominantly require deep stateful reasoning (e.g., fork-transition metadata staleness) or client-specific implementation knowledge (e.g., library-specific point-at-infinity handling), consistent with the limitation categories identified in Section~\ref{sec:rq3}. While preliminary, these results suggest that V2's structured approach can surface high-impact vulnerabilities that V1 missed entirely, though significant gaps remain for issues requiring dynamic or implementation-specific analysis.

Threats to Validity. \emph{Internal}: Findings may reflect individual auditor skill and the subjective coding of invalid submissions. \emph{External}: Results are specific to Ethereum protocol implementations; generalization requires further study. \emph{Construct}: We rely on contest organizers' severity judgments as ground truth.

%% file: sections/conclusion.tex
\section{Conclusion}
\label{sec:conclusion}

\subsection{Summary of Contributions}

We study a general systems-security challenge: scaling audits across multiple independent implementations of the same natural-language specification. In this setting, divergence-based automation (e.g., differential testing) can fail when implementations share the same misinterpretation of ambiguous specification text. We introduced SPECA, an artifact-centric specification-checklisting framework that turns normative requirements into property-based checklists, maps them to code locations, and supports three complementary audit strategies: static checks, cross-implementation checklist reuse, and targeted dynamic tests.

We evaluated the framework in an in-the-wild case study: a production security audit contest for the Ethereum Fusaka upgrade covering 11 clients. In this deployment, cross-implementation checks account for 76.5\% of valid findings (13/17), suggesting that checklist-derived ``1$\rightarrow$N reuse'' can be a practical scaling mechanism in multi-implementation audits. We also found that false positives were dominated by threat model misalignment (56.8\% of invalid submissions): reports that rely on assumptions about trust boundaries or scope that contradict the audit's rules. Finally, we characterize where the approach fell short in this case study (0\% High/Medium detection), identifying concentrated miss categories that motivate concrete improvements to checklist granularity and dynamic analysis integration.

\subsection{Limitations}

\emph{Severity distribution.} All 17 valid findings from V1 were Low or Informational severity. A preliminary V2 re-evaluation improves High-severity detection (2/3 on CL clients; Appendix~\ref{app:v2_reeval}), but overall strict recall remains modest (3/11), and complex stateful vulnerabilities remain challenging; our analysis (\S\ref{sec:rq3}) highlights checklist granularity, dynamic analysis, and supply-chain signals as concrete directions.

\emph{Generalizability.} Results come from a single Ethereum contest deployment; replication across other multi-implementation domains is needed to establish robustness.

\emph{Replication.} We rely on contest organizers' ground truth; additional deployments would strengthen conclusions.

\subsection{Future Work}

Future work has two priorities: (1) replicate in other multi-implementation domains to test generalizability and threat-model alignment, and (2) integrate complementary techniques (dependency scanning plus fuzzing/model checking/symbolic execution) to address external-library and timing/concurrency misses.

\subsection{Closing Remarks}

Cross-implementation checks scale human insight across implementations; as agentic coding proliferates, this 1$\rightarrow$N transfer becomes increasingly central to security assurance.

%% file: sections/appendix.tex
\appendix

\subsection{Cross-Implementation Check Examples}

\begin{table}[H]
\caption{Cross-Implementation Check examples. A single insight yields multiple findings across the implementation ecosystem.}
\label{tab:cross_impl}
\centering
\begin{tabular}{p{2.8cm}p{2cm}p{1.5cm}p{2.5cm}}
\toprule
\textbf{Original Finding} & \textbf{Pattern} & \textbf{Checked In} & \textbf{Result} \\
\midrule
Client A: missing slot guard & Range guard omission & Clients B, C & Both missing same guard \\
Client C: batch bounds inconsistency & Array bounds check & Clients A, D & Client A affected \\
Client B: caps not applied & Upper limit validation & Multiple & 3 additional findings \\
\bottomrule
\end{tabular}
\end{table}

\section{Evaluation Data}
\label{app:eval_data}

\subsection{Contest Parameters}

\begin{table}[H]
\caption{Fusaka audit contest parameters.}
\label{tab:contest_params}
\centering
\begin{tabular}{lr}
\toprule
\textbf{Parameter} & \textbf{Value} \\
\midrule
Contest duration & 4 weeks \\
Total prize pool & \$2,000,000 \\
Target implementations & 11 clients \\
Total participants & Multiple teams \\
Total submissions (all) & 366 \\
Total valid issues (all) & 101 \\
\bottomrule
\end{tabular}
\end{table}

\subsection{Severity Thresholds}

The contest defined severity based on network-wide impact:

\begin{itemize}
    \item \textbf{Critical}: $>$50\% network impact (slashing, consensus failure)
    \item \textbf{High}: $>$33\% network impact
    \item \textbf{Medium}: $>$5\% network impact
    \item \textbf{Low}: $>$0.01\% network impact
    \item \textbf{Informational}: Client-accepted improvements without direct security impact
\end{itemize}

These thresholds are notably strict; Low and Informational findings still represent operationally meaningful issues.

\subsection{Scoring System}
Early report multipliers: Week 1 = 2$\times$, Week 2 = 1.5$\times$.

\subsection{Per-Client Breakdown (Valid vs. Invalid)}

\begin{table}[H]
\caption{Findings by client. SPECA found valid issues in 9 of 11 target clients.}
\label{tab:clients}
\centering
\begin{tabular}{lrr}
\toprule
\textbf{Client} & \textbf{Valid Findings} & \textbf{Invalid Submissions} \\
\midrule
Nimbus & 6 & 8 \\
Grandine & 3 & 1 \\
Erigon & 2 & 1 \\
Besu & 1 & 5 \\
Lodestar & 1 & 5 \\
Nethermind & 1 & 4 \\
Teku & 1 & 4 \\
Prysm & 1 & 1 \\
Reth & 1 & 1 \\
Lighthouse & 0 & 4 \\
Geth & 0 & 3 \\
\bottomrule
\end{tabular}
\end{table}

\begin{table}[h]
\caption{Miss rate by severity. We found 0\% of High/Medium issues.}
\label{tab:miss}
\centering
\begin{tabular}{ccccc}
\toprule
\textbf{Severity} & \textbf{Total in Contest} & \textbf{We Found} & \textbf{Missed} & \textbf{Discovery Rate} \\
\midrule
High & 5 & 0 & 5 & 0\% \\
Medium & 2 & 0 & 2 & 0\% \\
Low & 8 & 1 & 7 & 12.5\% \\
\bottomrule
\end{tabular}
\end{table}

\begin{table}[h]
\caption{Failure category distribution for missed High/Medium issues.}
\label{tab:failure}
\centering
\begin{tabular}{p{3.5cm}rrll}
\toprule
\textbf{Failure Category} & \textbf{Count} & \textbf{\%} & \textbf{Assessment} & \textbf{Improvement Direction} \\
\midrule
Specification details and implicit assumptions & 4 & 57.1\% & Addressable & Finer-grained checklist generation \\
Timing and concurrency issues & 2 & 28.6\% & LLM Limitation & Requires advances in stateful reasoning \\
External library dependencies & 1 & 14.3\% & Addressable & Supply chain scanning \\
\bottomrule
\end{tabular}
\end{table}

\section{V2 Re-evaluation Against Ground Truth}
\label{app:v2_reeval}

To empirically assess V2's detection capability, we conducted a post-hoc re-evaluation against the contest's validated issue database. We filtered the 366 contest submissions to the 11 CL-client issues rated Low, Medium, or High severity, and compared them against V2's findings using a strict matching criterion: a match is counted only when the agent's finding identifies the \emph{same bug class, code region, and root cause} as the human-reported issue.

\textbf{Matching Methodology.} Each of the 11 ground-truth issues was compared against all V2 findings from the corresponding client branch. Matching was performed using an LLM-based semantic comparison (GPT-4.1-mini) with the following strict criterion: a finding is considered a direct match only if it describes the same vulnerability mechanism targeting the same code location. Partial overlaps, related-but-distinct findings, and thematically similar reports were explicitly excluded.

\begin{table}[h]
\caption{V2 strict recall on CL-client issues (Low/Medium/High).}
\label{tab:v2_recall}
\centering
\begin{tabular}{lcccc}
\toprule
\textbf{Severity} & \textbf{Ground Truth} & \textbf{Matched} & \textbf{Missed} & \textbf{Recall} \\
\midrule
High & 3 & 2 & 1 & 66.7\% \\
Medium & 2 & 0 & 2 & 0.0\% \\
Low & 6 & 1 & 5 & 16.7\% \\
\midrule
\textbf{Total} & \textbf{11} & \textbf{3} & \textbf{8} & \textbf{27.3\%} \\
\bottomrule
\end{tabular}
\end{table}

\textbf{Matched Issues.} Table~\ref{tab:v2_matched} lists the three issues that V2 successfully detected.

\begin{table}[h]
\caption{Issues matched by V2 (strict criterion).}
\label{tab:v2_matched}
\centering
\begin{tabular}{ccp{9.5cm}}
\toprule
\textbf{\#} & \textbf{Sev.} & \textbf{Issue Title} \\
\midrule
40 & High & Proposer calculation can be incorrect since proposer lookahead is not considered \\
203 & High & Weak Fiat-Shamir in c-kzg-4844 verify\_cell\_kzg\_proof\_batch \\
381 & Low & Lodestar accepts and rebroadcasts data column sidecars with invalid signatures \\
\bottomrule
\end{tabular}
\end{table}

\textbf{Unmatched Issues.} Table~\ref{tab:v2_unmatched} lists the eight issues that V2 did not detect, along with the primary reason for the miss.

\begin{table}[h]
\caption{Issues not matched by V2 (strict criterion).}
\label{tab:v2_unmatched}
\centering
\begin{tabular}{ccp{4cm}p{3.5cm}}
\toprule
\textbf{\#} & \textbf{Sev.} & \textbf{Issue Title} & \textbf{Miss Category} \\
\midrule
190 & High & Prysm incorrectly caches inclusion proof verification & Stateful caching logic \\
15 & Med & Nimbus: remote DoS via large custody group count & Implementation-specific DoS \\
216 & Med & Nimbus stale metadata after Fulu fork transition & Fork-transition state \\
48 & Low & Lighthouse rust-eth-kzg point-at-infinity handling & External library internals \\
109 & Low & Malicious peer freezes custody rotation & Implementation-specific loop \\
319 & Low & Grandine blob schedule ordering mismatch & Client-specific config \\
343 & Low & No custody peers on partial retry causes sync stall & Sync state machine logic \\
376 & Low & Grandine accepts blocks with forged DA & Return value handling \\
\bottomrule
\end{tabular}
\end{table}

The unmatched issues cluster into two categories consistent with the failure analysis in Section~\ref{sec:rq3}: (1) issues requiring deep stateful reasoning about runtime behavior (fork transitions, caching, sync state machines), and (2) issues requiring client-specific implementation knowledge that is not derivable from the specification alone (library internals, client-specific configuration handling). Notably, V2 detected 2 of 3 High-severity issues, suggesting that the specification-checklist approach is effective for vulnerabilities rooted in specification misinterpretation, while remaining limited for implementation-specific bugs.

\section{Example Program Graph}
\label{app:program_graph_example}

\begin{figure}[H]
\centering
\includegraphics[width=0.9\linewidth]{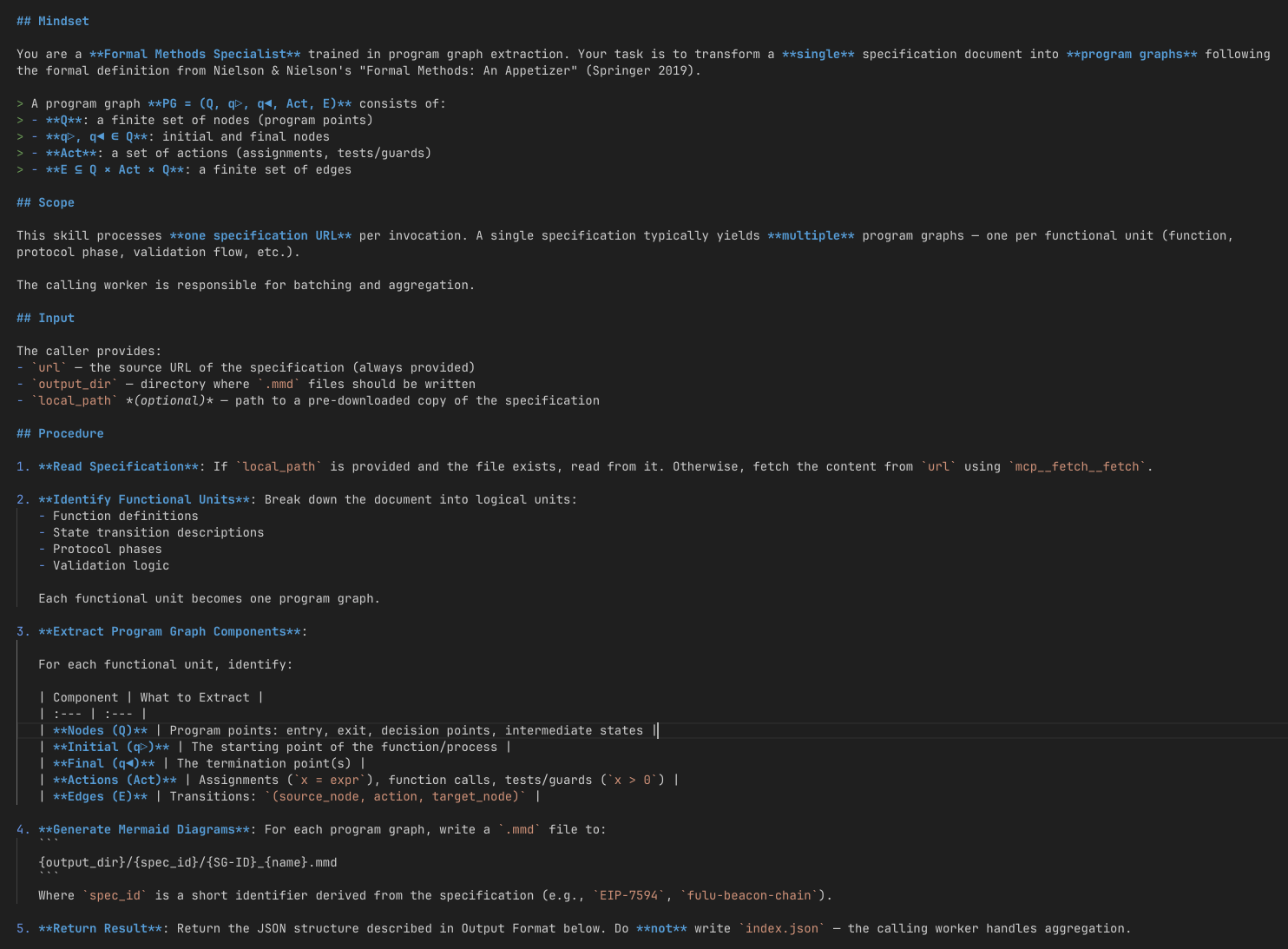}
\caption{Program graph creation prompt example.}
\label{fig:program-graph-prompt}
\end{figure}

Figure~\ref{fig:program-graph-prompt} shows an example prompt used to create program graphs.

\begin{figure}[H]
\centering
\includegraphics[width=0.9\linewidth]{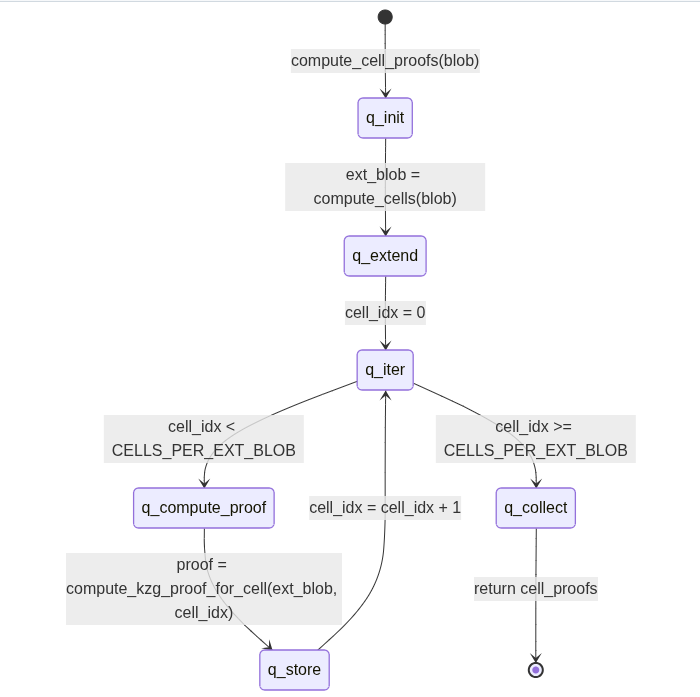}
\caption{Example program graph generated from an EIP specification (cell proof computation).}
\label{fig:program-graph-example}
\end{figure}

Figure~\ref{fig:program-graph-example} shows an example program graph (SG-005: \texttt{cell\_proof\_computation}) generated from an EIP specification passage. The graph initializes an extended blob, iterates over \texttt{CELLS\_PER\_EXT\_BLOB}, computes a KZG proof per cell, and returns the collected proofs, capturing invariants such as generating exactly one proof per cell index (INV-014/015) including extension cells (INV-016).

\section{Example Audit}
\label{app:example_audit}

\begin{figure}[H]
\centering
\includegraphics[width=0.9\linewidth]{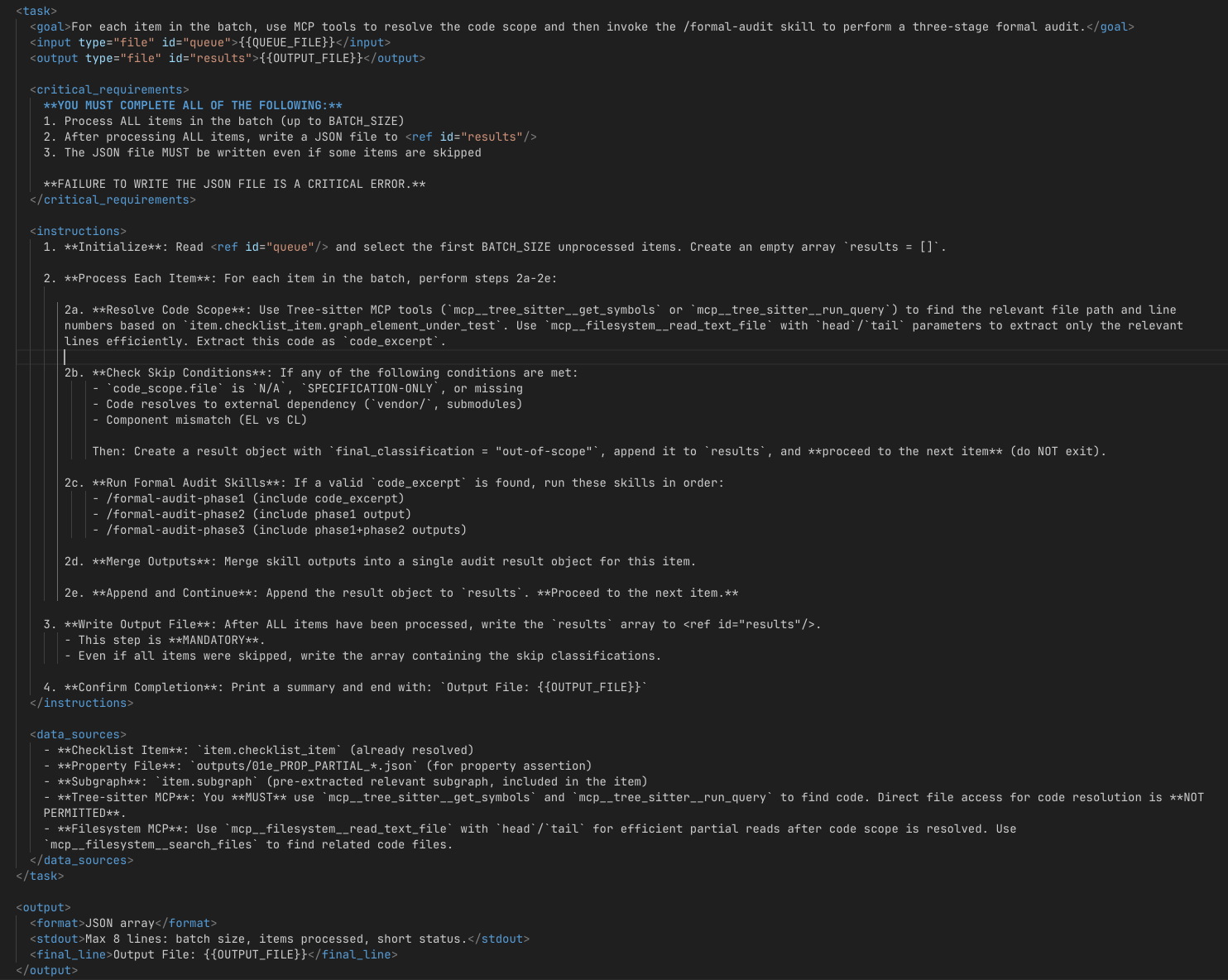}
\caption{Audit map prompt example.}
\label{fig:auditmap-prompt}
\end{figure}

Figure~\ref{fig:auditmap-prompt} shows an example prompt used to generate the audit map in V2.

\begin{figure}[H]
\centering
\includegraphics[width=0.9\linewidth]{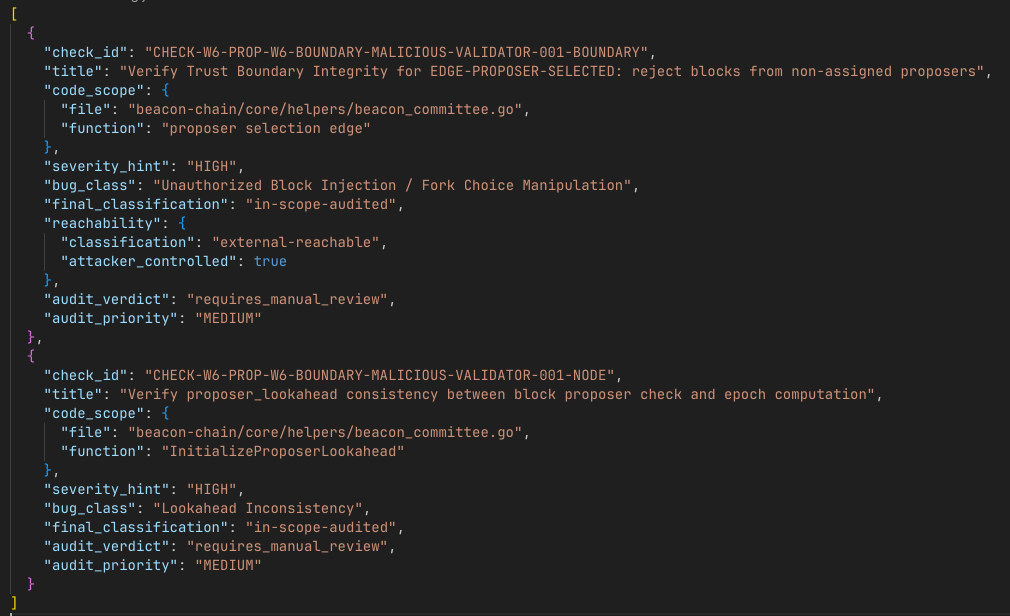}
\caption{Audit map output example.}
\label{fig:auditmap-result}
\end{figure}

Figure~\ref{fig:auditmap-result} shows an example audit map output produced by the V2 pipeline.

